\documentclass[floatfix,aps,amsmath,nofootinbib,10pt]{revtex4}
\usepackage{listings}
\usepackage{graphicx}
\usepackage{bm}
\usepackage{rotating}
\usepackage{array}
\usepackage{amsmath}
\usepackage{amssymb} 
\usepackage{mathrsfs} 
\usepackage{cancel}
\usepackage{framed,multirow}

\usepackage{latexsym}
\usepackage{epstopdf }
\usepackage{url}
\usepackage{xcolor}
\usepackage{color}
\definecolor{newcolor}{rgb}{.8,.349,.1}
\usepackage{hyperref} 
\usepackage{cleveref} 
\hypersetup{colorlinks = true, 
	    linkcolor = black, 
	    urlcolor = blue,
            citecolor = blue} 
\usepackage[doipre={doi:~}]{uri}

\def\({\left(}
\def\){\right)}
\def\[{\left[}
\def\]{\right]}

\def\e{\begin{equation}}
\def\q{\end{equation}}
\def\m{\begin{eqnarray}}
\def\n{\end{eqnarray}}

\begin{document}

\date{\today}% It is always \today, today,

\title{Doppler Tracking Data of Martian Mission Tianwen-I and Upper Limit of Stochastic Gravitational Wave Background}
%\tnotetext[tnote1]{This is an example for title footnote coding.}

\author{Xiaoming Bi$^a$}

\author{Zhongkai Guo$^b$}
\author{ Xiaobo Zou$^a$}
\author{Yong Huang$^c$}
\author{Peijia Li$^c$}
\author{Jianfeng Cao$^d$}
\author{Lue Chen$^d$}
\author{Wenlin Tang$^{e}$}
\thanks{Corresponding author: tangwenlin@nssc.ac.cn }
 \author{Yun Kau Lau$^{f}$}
\thanks{Corresponding author: lau@amss.ac.cn }

\affiliation{$^a$School of Physical Science and Technology, Lanzhou University, Lanzhou 730000, China;}
\affiliation{$^b$Beijing Institute of Space Mechanics and Electricity, China Academy of Space Technology, Beijing 100094, China;}
\affiliation{$^c$Shanghai Astronomical Observatory, Chinese Academy of Sciences, Shanghai 200030, China; 
    Key Laboratory of Planetary Sciences, Shanghai Astronomical Observatory, Chinese Academy of Sciences, Shanghai 200030, China;}
\affiliation{$^d$Key Laboratory of Electronics and Information Technology for Space System, National Space Science Center, Chinese Academy of Sciences, Beijing 100190, China;}
\affiliation{$^e$Science and Technology on Aerospace Flight Dynamics Laboratory, Beijing Aerospace Control Center, Beijing, China; }
\affiliation{$^f$Institute of Applied Mathematics, Morningside Center of Mathematics, LSSC, Academy of Mathematics and System Science, Chinese Academy of Sciences, 55, Zhongguancun Donglu, Beijing, 100190, China.}

\begin{abstract}
Two way ranging data for spacecraft  tracking of  China's first Martian mission Tianwen-I is analysed. Shortly before the spacecraft entered the Mars parking orbit, the two way coherent microwave link between the spacecraft and the Earth resembles a long arm gravitational wave interferometer, with both the spacecraft and the Earth regarded as in an approximate free falling state.  By carefully selecting and analysing data segments of the time series of the two way ranging data during this time span, a parametric statistical model is built for the data segments and an upper limit for the stochastic gravitational waves background (SGWB) is then estimated within the frequency window 0.1Hz to 0.1 mHz. The upper bound  improves considerably on those obtained before. In particular, around the deci-Hz band, there is a three orders improvement on the bound obtained previously by the two way ranging data of the  Chang e 3 mission. Scientific applications of the upper bound is then considered and a weak upper bound is worked out for axions  which is a promising candidate for ultra light dark matter.

\end{abstract}
\maketitle

%% For linenumbers
%\linenumbers
\section{Introduction}

Tianwen-I is China's first Mars exploration mission and it heralds the dawn of a new era of deep space exploration for China. It was launched  on July 23, 2020
from the Wenchang Space launch center in Hainan province and arrived on Mars in May 1, 2021.
During its seven months interplanetary journey to Mars, as part of the communication link between the spacecraft and the ground system of the  Chinese deep space network, the spacecraft was tracked by means of two way coherent  X band ranging by both the Jiamusi and Kashi ground stations. 

\begin{figure}[htb]
    \centering
    \includegraphics[width=0.7\textwidth]{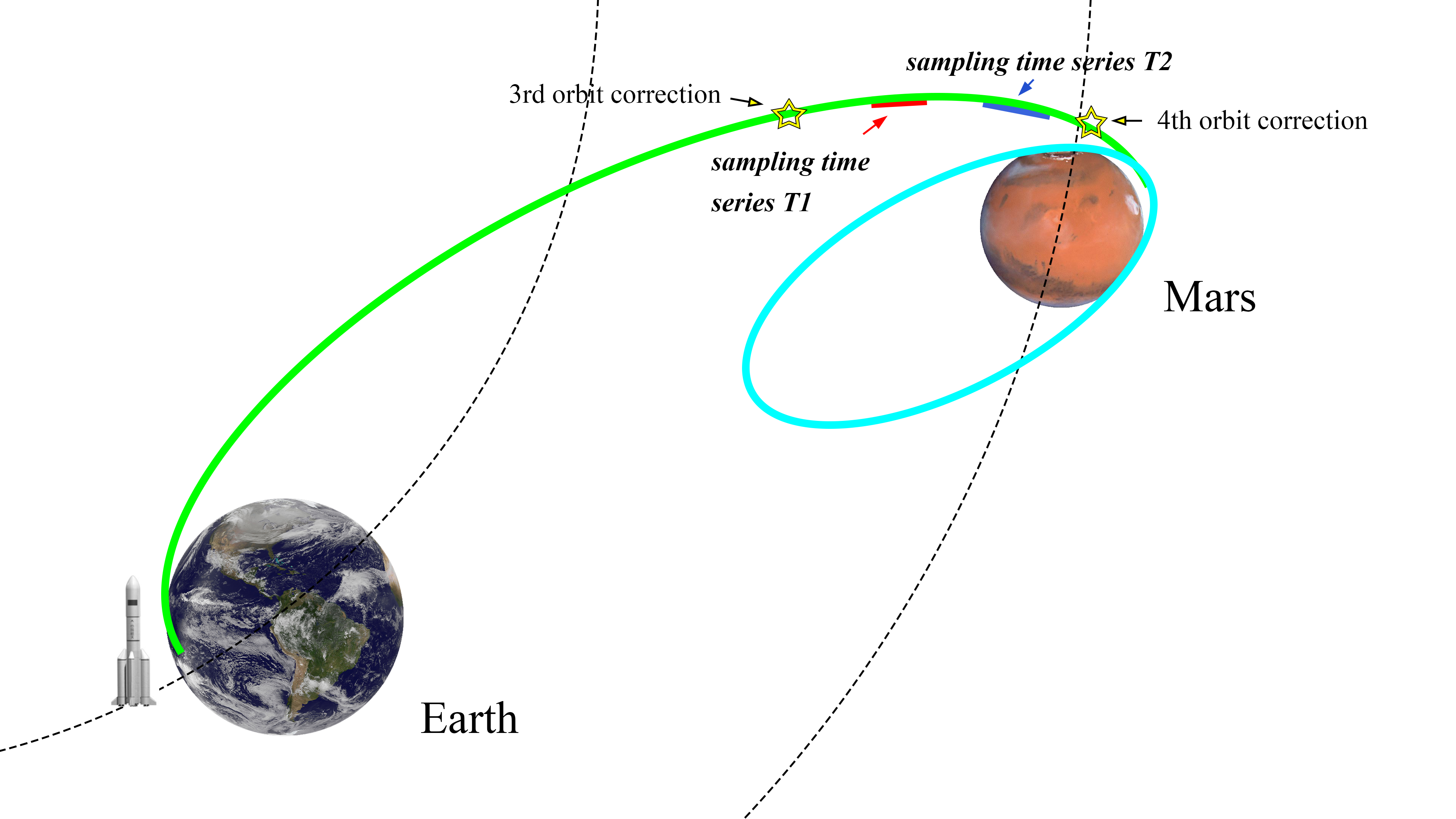} 
    \caption{The selected date segment $T_1$ (red line) is on January 8, 2020 and $T_2$ (blue line) is on February 5, 2020. Both dates fall between the 3rd and 4th orbit correction. The green line, cyan line and dashed line represent the Earth-Mars transfer orbit, capture orbit and planetary motion trajectory, respectively.}
    \label{Fig.1}
\end{figure}

The transponder scheme in tracking the range rate observable means that the two-way microwave link between the Earth and the spacecraft 
resembles a single-arm interferometer whose arm-length is at a staggering $ 10^8$ km, on a much longer length scale than that for a spaceborne gravitational wave detector \cite{yang2022orbit}, though at least at the current level of sensitivity, the measurement precision is only at the level of 0.1 mm/s. Furthermore, after the third orbit maneuver and before the fourth orbit maneuver when the  spacecraft is about to enter  into the Mars parking orbit, the spacecraft may be in practice regarded as a free-falling body subject only to gravitational influence in empty space, with non-gravitational force and planetary gravity perturbation as residual acceleration noise. With the Earth also regarded as a free-falling body at the same time, the two-way microwave link provided interferometric measurement of the Doppler shift between two free-falling bodies, as in the case of gravitational wave detection (see \cref{Fig.1}). 

As part of the scientific applications of the tracking data of the Tianwen-I  mission, it is  natural to ask whether the range rate data may be employed to provide useful information concerning gravitational wave (GW) physics and experimental test of general relativity. 
Built upon the expertise accumulated from the Chang'e-3 lunar mission \cite{tang2017chang}, the present work aims to work out an upper bound on the cosmological background of gravitational waves, at a frequency band much boarder than that provided by the Chang'e-3 tracking data and at a much improved level of precision. A scientific application of this upper bound to constrain ultra light dark matter is  also presented. 
    
The paper is structured as follows. Section 2 describes the detailed data analysis process. Section 3 focuses on the  estimation of the upper limit of SGWB. A scientific application of this upper limit to constrain ultra-light dark matter is presented in section 4. Some brief remarks concerning future work are provided in section 5 to conclude the paper. 
\label{sec1}

\section{Data analysis}

In this section, we will present the details of the data analysis for the Tianwen-I two-way range rate data. The data processing can be divided into the following steps: 1. Select data segments suitable for constraining SGWB. 2. Smooth the data for 9 seconds to suppress random noise. 3. Utilize the Autoregressive moving average (ARMA) model to obtain data statistical parameters and estimate PSD. 4. Constrain SGWB within the 0.1 mHz-0.1 Hz frequency band.

\subsection{ Data structure and selection.}

We will consider the  two way ranging data measured by the Jiamusi station for the Tianwen-1 mission. The position of the spacecraft are recorded in both the geocentric and heliocentric frames, from which the round trip light time can be inferred with $T_1=917.15 \mathrm{s}$ and $T_2=1206.62 \mathrm{s}$. During the interplanetary journey, everyday a segment of data was taken for 
a period of 2-4 hours. This generates a  discontinuous segments of time series of data that span over a period of seven months with 1s sampling time. Due to the presence of bias originated from unknown factors, the velocity of the measured for the spacecraft is generally different for each segment and in addition the data gap in between two consecutive data segments. As down time is much longer than the sampling time, it does not seem possible to piece together the data segments to form a coherent data set that covers a long period of time. Instead we need to work on each individual segment of time series of data and a careful selection of data segments becomes the first task. 

In our data selection, major factors such as the round trip time, noise level of the data series, and the influence of planetary gravity on the Doppler data are taken into account. We choose two data segments between the third correction of the spacecraft's orbit and the fourth correction prior to entering the Mars orbit. The two selected data segments  are from 11:03:37.5 to 12:43:22.5 UTC on January 8, 2020 and 03:33:32.5 to 06:33:31.5 on February 5. (see \cref{Fig.1} for a sample of the time series of data). The spacecraft experienced a state of free and stable flight during these two selected time intervals.

\begin{figure}[htb]
    \centering
    \includegraphics[width=.5\textwidth]{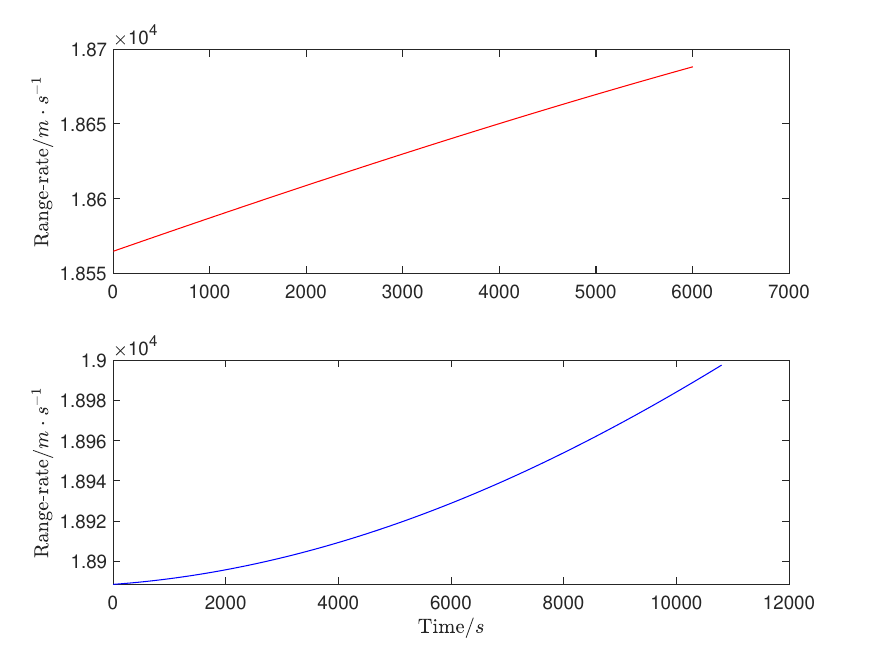} 
    \caption{Two typical Doppler time series $T_1 \& T_2$. The spacecraft may be regarded as a free falling body during these measurement and control stages.}
    \label{Fig.2}
\end{figure}

\begin{figure}[htb]
    \centering
    \includegraphics[width=.5\textwidth]{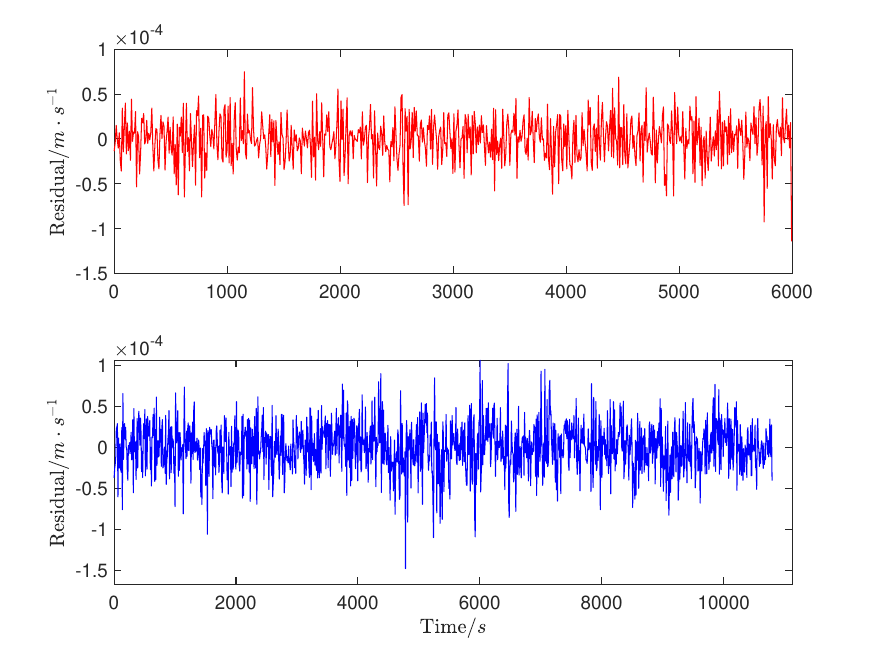} 
    \caption{Residuals of two Doppler time series $T_1 \& T_2$. The standard deviation is $2.5\times10^{-5}m/s$ for the red line and $3.1\times 10^{-5} m/s$ for the blue line.}
    \label{Fig.3}
\end{figure}

\subsection{Data processing. }

The initial data processing for the Tianwen-I mission is similar to that of Chang'e-3 \cite{tang2017chang}. From the establishment of a polynomial model for spacecraft orbits, the residual sequence is stated as follows.
\begin{equation} 
\epsilon(t_i)=\rm v_{Doppler}(t_i)-v_{orbit}(t_i)  .
\end{equation}
where $\rm v_{orbit}(t_i)$ is the modelled velocity of the spacecraft,$\rm v_{Doppler}(t_i)$ is the time series of velocity data. In this work, we use 5th-order Chebyshev fitting to process orbit speed. The RMS of the speed residual is $2.543\times 10^{-5} \mathrm{m/s}$, which is comparable to the precision of Tianwen-I's speed measurement. The impact of time delay induced by the gravitational field can be safely neglected when considering the sampling time. In order to suppress random fluctuations in the data, we smoothed the two-way range rate data for 9 seconds.

In the next step, instead of using the standard Fourier approach, we will build a parametric statistical model for the data time series, based on the weak stationary structure of the time series to be verified in what follows. With ARMA processes, stationary time series can be modeled, involves the following steps:\cite{chan2008time}:

\vskip 10pt
\par\noindent\textbf{Data Randomization \& Stationarity test}: 

In general, an ARMA $(p,q)$ time series may be written as
\begin{equation} Y_{t}=\phi_{1} Y_{t-1}+\cdots+\phi_{p} Y_{t-p}+e_{t}-\theta_{1} e_{t-1}-\cdots-\theta_{q} e_{t-q} ,  
\label{eq1}
\end{equation}
where  $Y_t=\{y_i\}_t$ is the observed time series, \{$\phi_i$\} and \{$\theta_i$\} are the coefficients of AR and MA process. The ACF of time series with time lag $k$, 
\begin{equation}
\begin{array}{cc}
     & \mathrm{ACF}(k)=\dfrac{c_k}{c_0} ,\\
     & c_k=\frac{1}{T-1}\sum_{t=1}^{T-k}(y_t-\bar{y})(y_{t+k}-\bar{y}) ,
\end{array},
\end{equation}
encodes information of periodic signals in time series.
%\textcolor{red}{encodes} 
\begin{figure}[htb]
    \centering
    \includegraphics[width=.7\textwidth]{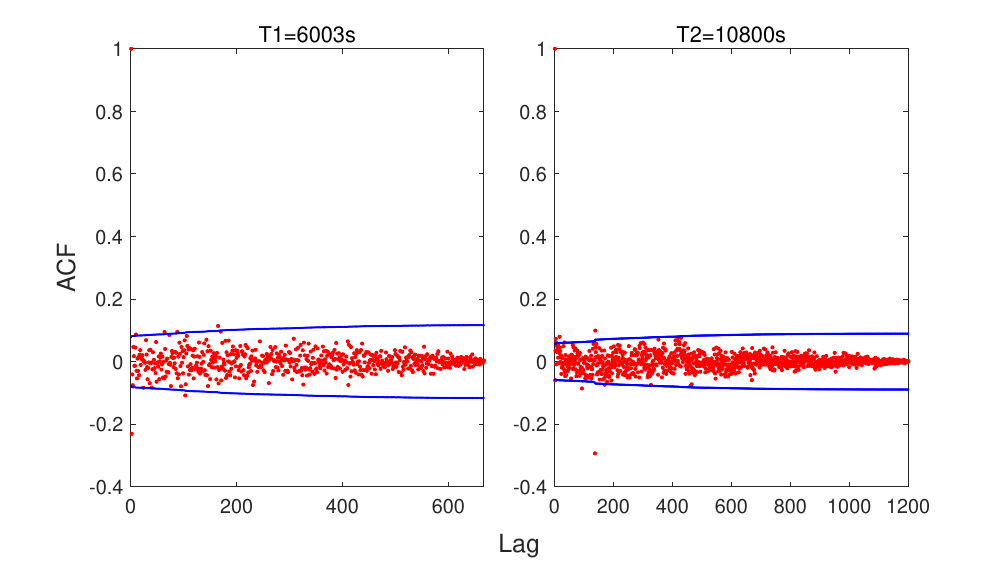} 
    \caption{Autocorrelation coefficients function of the residual series. The blue dots represents 2$\sigma$ bounds in estimation. }
    \label{Fig.4}
\end{figure}
The ACFs of $T_1 \& T_2$ are shown in \cref{Fig.4}, as two independent constraints in two different time  for upper limits of $\Omega_{GW}$ in \cref{Fig.7}. However, in the time series with radical constraints, ACF does not strictly meet the estimation bias of truncation to $2 \sigma$, and exhibits a positive correlation similar to that in the Cassini task (see fig.2 in \cite{armstrong2003stochastic}). These significant deviations beyond the estimation error can prove the existence of the periodic noise in Doppler signal. Compared with the ACF of $T_1$ and $T_2$, $T_1$ has fewer points with significant autocorrelation, resulting in fewer model parameters and overfitting.

Their significant differences in the low-frequency band of $\Omega_{GW}$ stem from differences in arm length(round-trip light time) and model parameters.  \cref{Fig.3} suggests that the data is not a completely random (white noise) sequence. Smaller ARMA orders can be used to fit the power spectral density (PSD), but this may not be universally applicable across the entire data segment, due to the interference from  non-gravitational forces in deep space. 

Another interesting feature of data segments is their weak stability. Strict stationarity implies that for any time $t$ and time lag $k$ $(k < t)$, time series $Y_t$ and $Y_{t-k}$ have the same joint probability, ensuring that the statistical properties of the data do not change significantly over time. However, strict stationarity is often not achievable in sampled data. Therefore, we typically assess the stationarity of lower-order moments as a test, such as second-order moment stationarity, which is described as follows:
\begin{itemize}
    \item The mean function is constant over time. 
    \item The covariance of any two time series on the sampled data is only related to the time lag $k$: ${\rm Cov}(Y_t,Y_{t-k})= {\rm Cov}(Y_0,Y_{k})$.
\end{itemize} 
The Augmented Dickey–Fuller (ADF) \cite{dickey1979distribution} test and the Kwiatkowski-Phillips-Schmidt-Shin (KPSS) \cite{kwiatkowski1992testing} test are commonly utilized to assess the stationary structure of the residual sequence. We adopt $\alpha=5$ (significance levels) for our hypothesis-testing. The results of both tests were consistent with our expectations, combined with ACF in \cref{Fig.4} indicating that the two residual series are indeed weakly stationary.

\vskip 10pt
\par\noindent\textbf{Model identification}: 

The parameters $(p)$ and $(q)$ determine the shape of the power spectrum. We identify the appropriate ARMA model that best fits the data based on the corrected Akaike’s Information Criterion $(AIC_c)$ to avoid overfitting. In the previous analysis of the Chang'e 3 Doppler data, the model was ARMA(0,1). While searching the entire Tianwen-1 data stream, a few time series exhibit significant correlations, and there are few points outside the $(2\sigma)$ lines in the time lag, as shown in \cref{Fig.4}, compared with the ACF in the Chang'e 3 mission \cite{tang2017chang}. The significant lag correlation observed in Cassini tasks is attributed to unmodeled noise \cite{armstrong2003stochastic}. Therefore, we anticipate that the long baseline of the Tianwen-I mission TT\&C link introduces more complex noise sources, such as space plasma with scintillation characteristics.

In our task, $T_2$ exhibits more significant ACFs that exceed the $(2\sigma)$ estimation error lines, suggesting that lower-order ARMA models may not be accurate enough compared to $T_1$. The model for Tianwen-1 Doppler data is ARMA$(1,2)$ for $T_1$ and ARMA$(0,2)$ for $T_2$, respectively.

\begin{figure}[htbp]
	\centering
	\includegraphics[width=0.7\textwidth]{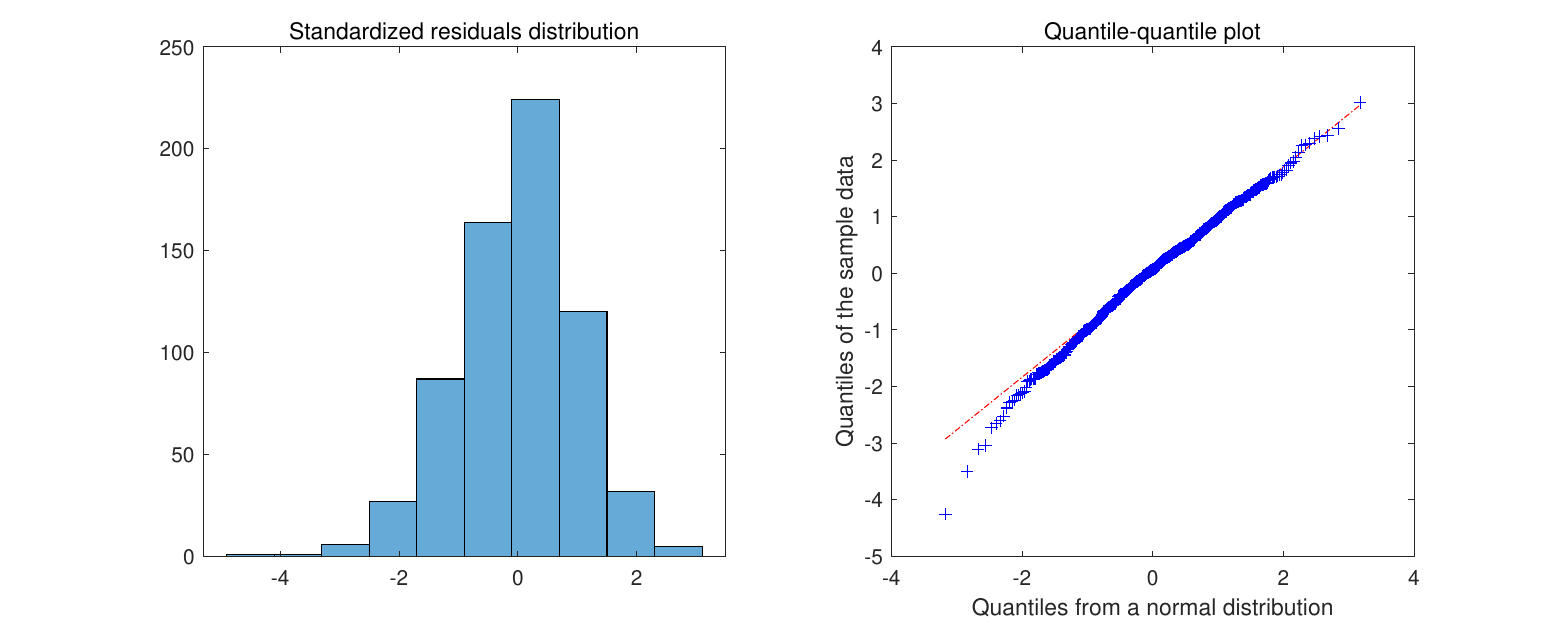} \\
	\includegraphics[width=0.7\textwidth]{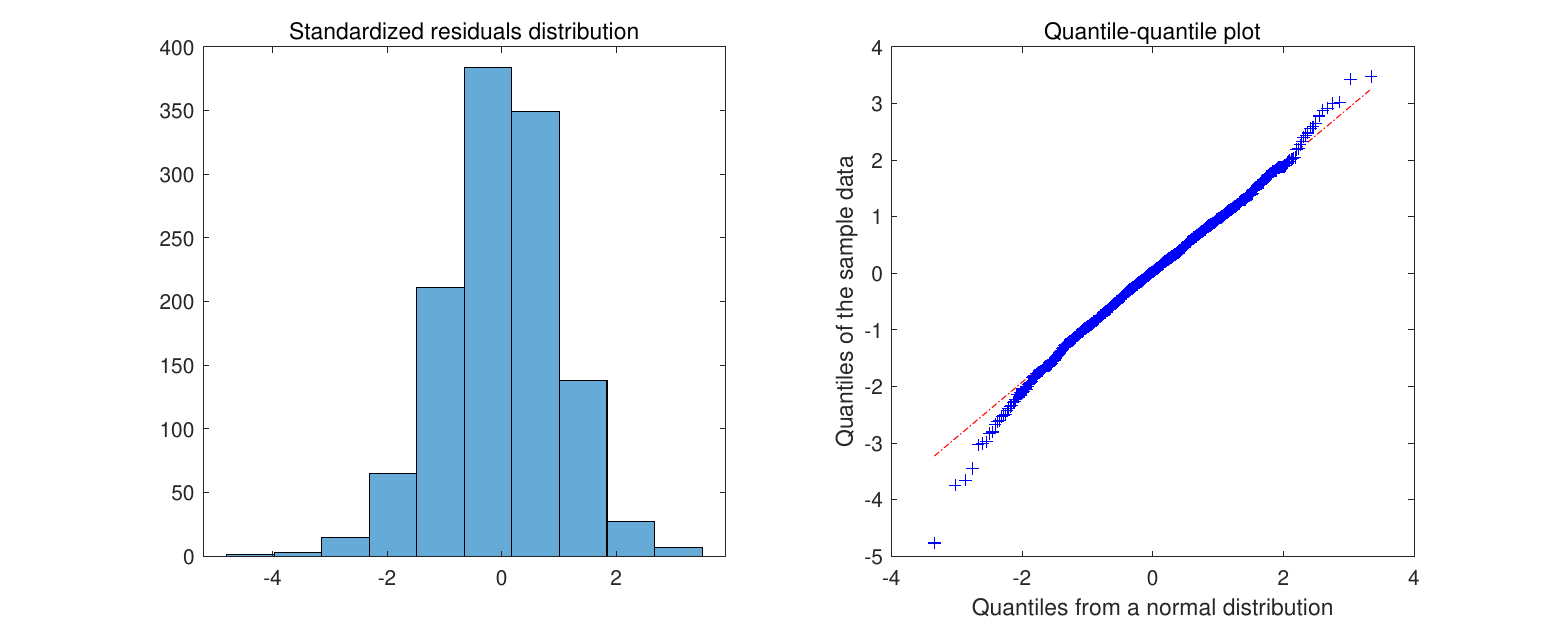}
	\caption{Model diagnostic for $T_1$ (top) and $T_2$ (bottom) time series. Left: quantity distribution of normalized residuals. Right: Quantile-Quantile plot of residual distribution and Normal distribution.}
	\label{Fig.5}
\end{figure}
\vskip 10pt
\par\noindent\textbf{Model diagnostics}: 

By examining whether the residuals between the estimated and actual values adhere to a Gaussian distribution, we are able to  assess if the model successfully captures spectral information. An accurate model makes the residuals closer to white noise. If the residuals from the model pass the Ljung-Box test \cite{ljung1978measure}, we can be sure that the model quality is acceptable. \cref{Fig.5}  displays the approximate Gaussian structure of the residuals between the estimated and actual values. 

\subsection{The upper limits of SGWB.}

Having established the statistical model for the time series of Tianwen-1 data, we can now calculate the power spectrum of the residuals to estimate the magnitude of the SBGW. We then map the velocity power spectrum to the characteristic root-mean-square of the strain and work out the upper bound of the PSD of SBGW.

Based on the minimum $AIC_c$ \cite{akaike1973maximum}, the order $p, q$, and corresponding parameter series, \{$\phi_i$\} and \{$\theta_i$\} in \cref{eq1} may be inferred. For time series $T_1$ and $T_2$, this is an independent process, wherein the inference result of $T_1$ is

\begin{equation} \begin{array}{cc}
     & \theta_{1}=-0.2795,\theta_{2}=-0.0670, \\
     & \sigma_{T1}^2=6.001\times10^{-10} ,
     \end{array}
\end{equation}
and for $T_2$ is
\begin{equation} \begin{array}{cc}
     & \phi_{1}=0.9308,\theta_{1}=-1.0065,\theta_{2}=0.1156, \\
     & \sigma_{T2}^2=9.420\times10^{-10} .
     \end{array}
\end{equation}

These parameters  determine the shape of the power spectrum. However, the spectra is still not yet the final spectrum for SGWB. In addition, considering that the significant autocorrelation in \cref{Fig.4} has not been modeled yet,  the spectrum of GW would still be buried under noise as $S_{GW}(f) < S(f)$. The estimated error of the autocorrelation coefficients function in \cref{Fig.4} is \cite{chan2008time}

\begin{equation} \sigma(\rho_{k})=\sqrt{\frac{1}{n}\left[1+2 \sum_{j=1}^{k} \rho_{j}^{2}\right]},\quad  k>q  
\end{equation}

The PSD of \{$Y_i$\} is

\begin{equation} S_{\mathrm{DS}}(f)=\frac{\sigma^{2}}{2 \pi}\left|\frac{1+\sum_{\alpha=1}^{\alpha_{max}}\theta_{\alpha} e^{i2\alpha\pi  f}}{1+\sum_{\beta=1}^{\beta_{max}}\phi_{\beta} e^{i2\beta\pi  f}}\right|  \end{equation}

In radio link observations, the presence of gravitational waves (GWs)  will generate gravitational redshift. In particular, for a weak plane GW that is oriented in a fixed direction, the redshift of the signal may be modelled as a three-pulse response \cite{estabrook1975response}, 
\begin{align}
    \frac{\Delta \nu}{\nu}=\frac{1}{2}(1+\alpha)G\sin\omega\left(t+t_{0}-\frac{1}{2} T\right)  \nonumber
    -\alpha G \sin \omega\left(t+t_{0}- \frac{1}{2} \alpha T\right)\nonumber 
    -\frac{1}{2}(1-\alpha) G \sin \omega\left(t+t_{0}+\frac{1}{2} T\right) ,
\end{align}
where G is the Fourier component of the amplitude of the particular circular frequency $\omega=2\pi f$, $\alpha=\hat{k}\cdot \hat{n}$ is the cosine angle between the gravitational wave (GW) and the radio link, and $T$ is the round trip time of the observed photon.

Given the transfer function $G(f)$, we may  associate the polarization-averaged energy of the isotropic GW background with the frequency shift power spectral density (PSD) described by
\begin{align}
\Big(\frac{\bar\Delta \nu}{\nu}\Big)^2&=\frac{1}{2}G(f)^2[ 1 - \frac{1}{3}\cos \left( {2\pi f{T_2}} \right) \nonumber 
- \frac{{\left( {3 + \cos \left( {2\pi f{T_2}} \right)} \right)}}{{{{\left( {\pi f{T_2}} \right)}^2}}} + \frac{{2\sin \left( {2\pi f{T_2}} \right)}}{{{{\left( {\pi f{T_2}} \right)}^3}}}]\nonumber \\
&=\frac{1}{2}G(f)^2 \bar{R_2}(f) ,
\end{align}
which is related to the spectrum of Doppler-shift of the spacecraft-station link, $S_{DS} $, given by \cite{Maggiore:1999vm}
\begin{equation}
    S^{\mathrm{gw}}(f)=\bar{R}_{2}(f) S_{h}(f) \leq S_{DS}(f),
\end{equation}
where $ S_h $ is the characteristic rms strain related to the amplitude and $ S_{DS} $ is the spectrum of the Doppler time series.
The dimensionless energy density of GW defined as
\begin{equation} \Omega_{\mathrm{GW}}(f)=\frac{8 \pi^{2} f^{3}}{3 H_{0}^{2}} S_{h}(f) \leq \frac{8 \pi^{2} f^{3}}{3 H_{0}^{2}} \frac{S_{DS}}{\bar{R_2}}  \end{equation}

\begin{figure}[htb]
    \centering
    \includegraphics[width=.7\textwidth]{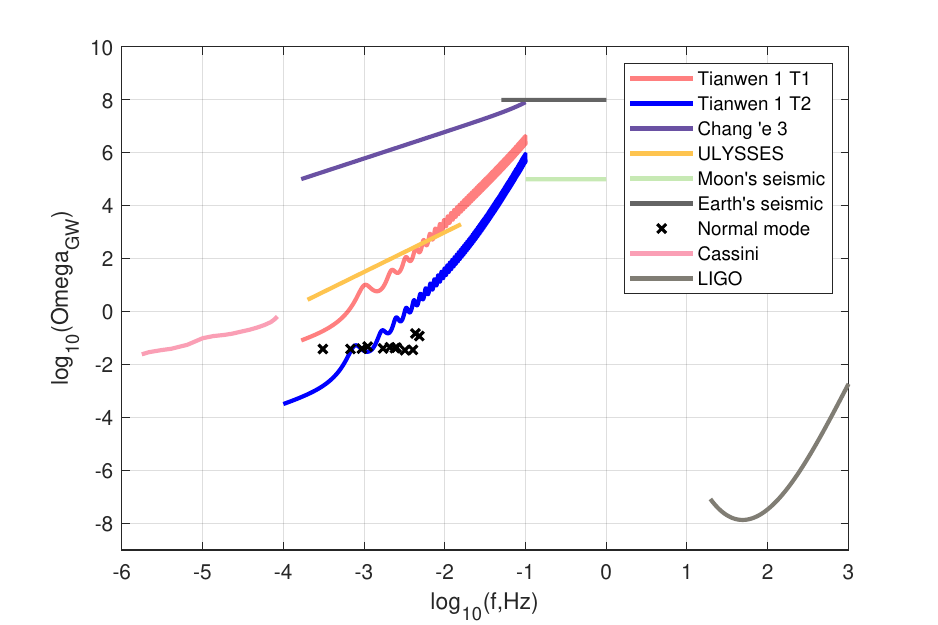} 
    \caption{The conservative (red line) and radicalupper(blue line) upper limits of SGWB compare with Chang'e 3 mission \cite{tang2017chang}, ULYSSES \cite{bertotti1995search}, Lunar seismic \cite{coughlin2014upper}, Earth seismic \cite{coughlin2014constraining}, normal modes \cite{coughlin2014constraining2}, Cassini spacecraft \cite{armstrong2003stochastic} and LIGO \cite{LIGOScientific:2019vic}.}
    \label{Fig.6}
\end{figure}

\begin{figure}[htb]
    \centering
    \includegraphics[width=.7\textwidth]{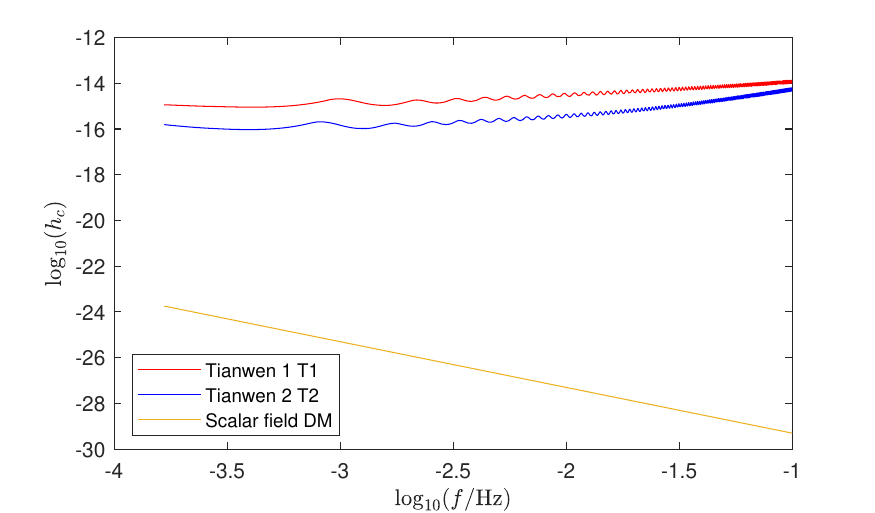} 
    \caption{Constains on $h_c$ when SGWB only from the scalar dark matter between detectors \cite{Khmelnitsky:2013lxt}, with the dark matter density $\rho_{\mathrm{DM}}=0.3  \mathrm{GeV/cm^3}$.}
    \label{Fig.7}
\end{figure}

The latest constraints on SGWB upper limits are shown in \cref{Fig.6}. The results of the two Tianwen-1 time series are depicted by red and blue lines. Due primarily to the long baseline of the two-way microwave link between the spacecraft and the deep space network, Tianwen-I mission improves on the upper bound on SGWB set by the  ULYSSES mission in the 0.1 mHz-0.1 Hz frequency band. 

\section{Scientific applications}

We shall next study the scientific applications of this bound in our understanding of dark matter (DM) in cosmology. 

Standard Lambda cold Dark Matter ($\mathrm{\Lambda CDM}$) cosmological model is very successful in describing the large-scale structure of the universe.
The possibility of dark matter existence can be demonstrated through galaxy rotation curves, evolution of large-scale structures, and gravitational lensing observations.
Ultra-light scalar fields are hypothetical particles with extremely low masses, potentially constituting a significant component of dark matter \cite{Hui:2016ltb}.  These fields, often theorized in extensions of the Standard Model of particle physics and ingeniously designed to address the strong CP problem in QCD, have masses in the range of $10^{-22}$to $10^{-10} \mathrm{eV}$, which are called axions or ultra-light bosons. 

The SGWB is a superposition of GWs that originated from numerous independent and unresolved astrophysical sources, creating a continuous background noise. This background  contains a wealth of information concerning the early universe and fundamental physical processes. The linkage between ultralight bosons and SGWB is twofold. Firstly, ultralight bosons, as possible candidate for cold dark matter, may generate observable SGWB through photons redshift \cite{Khmelnitsky:2013lxt}, \cite{Kim:2023pkx}. This may be well described by the scalar field exhibiting time-oscillating pressure within the galactic halo and this generates oscillations in the gravitational potential, and is being characterized as monochromatic waves with the frequency equal to the mass of the particles. The GWB generated by this effect is represented by:

\begin{equation}
h_{c}= \sqrt{15} \Psi_{c}=2 \times 10^{-15}\left(\frac{\rho_{D M}}{0.3 \mathrm{GeV} / \mathrm{cm}^{3}}\right)\left(\frac{10^{-23} \mathrm{eV}}{m}\right)^{2}
\end{equation}

In \cref{Fig.6}, We have displayed the sensitivity of Tianwen-I and ULDM cosmological processes in the current frequency band and the gap between them. 

\begin{figure}[htb]
    \centering
    \includegraphics[width=.7\textwidth]{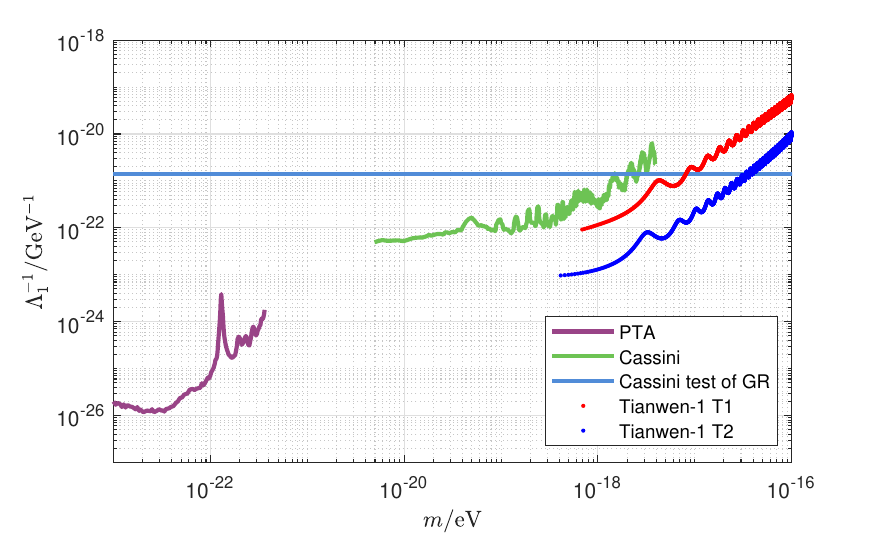} 
    \caption{Upper limits of linear coupling $\Lambda_1^{-1}$ bewteen ULDM and matter \cite{Blas:2016ddr}, compared with PTA \cite{Porayko:2014rfa}, Cassini SGWB \cite{armstrong2003stochastic} and Cassini test of GR \cite{bertotti2003test} .}
    \label{Fig.9}
\end{figure}

Secondly, ultra-light bosons can form clouds around stellar objects like for instance black holes \cite{Brito:2017wnc}, in which low-energy bosons near spinning black holes can trigger superradiant instability when the boson frequency $\omega_R$ satisfies the superradiant condition $0 < \omega_R < m\Omega_H$, where $\Omega_H$ is the horizon angular velocity and m is an azimuthal quantum number. The energy-momentum tensor of  axions perturbs the metric of binary systems and is described phenomenologically by a scalar field. The scalar field generates extra tidal force that contributes to the SGWB. Further,  axions also undergo weak coupling with matter, and the upper bound of SGWB imposes a constraint on its coupling strength. At the energy scale of $10^{-22}\mathrm{eV} $, equivalent to the frequency regime around $10^{-9}$ Hz, the  Pulsar Timing Array (PTA) \cite{Blas:2016ddr}, \cite{Blas:2019hxz} gives an upper limit on the coupling strength. The upper bound of SGWB  deduced in the preceding section imposes an upper limit on the first-order coupling coefficient $\Lambda_1$ of ULDM at the energy regime around $10^{-22}$ to $10^{-18} \mathrm{eV}$, 
corresponding to the frequency window 0.1mHz to 0.1 Hz (see \cref{Fig.9}).

\section{Concluding remarks}

Tianwen-I was the first Chinese mission to go beyond the Earth-Moon system and went into deep space. In anticipation of more Chinese deep space missions in the future, a window will be open to conduct experimental tests of general relativity in space and detection of gravitational waves in particular. It is also expected that in the coming years both the precision and quality of spacecraft Doppler tracking will be improved. For instance, dual-band ranging or the employment of Ka-band will be obvious ways to improve the data precision. From the noise analysis of the ranging data, time delay generated by the propagation of signals through the Earth's troposphere and ionosphere(\cite{Li:2008}) will be major noise sources. 
With a permanent lunar station in planning in both China and the US, future antennae for DSN may be partly built on the lunar surface instead of on Earth, the Earth atmosphere-related noise sources may then be bypassed and the ranging accuracy will have vast potential for improvement. A Doppler data of $\mathrm{nm/s}$ accuracy may not be out of reach and the prospect of detecting gravitational waves from astrophysical sources like supermassive black hole binaries (\cite{thorne1976gravitational}) will then be a realistic possibility. We hope the present work will be a small step to build towards this promised land. 

\section{Acknowledgments}
The work is supported by the National Key Research and Development Program of China under Grant 2021YFC2202501. The work of Wenlin Tang is supported by  National Natural Science Foundation of China (Grant No.~11905244). The work of Yong Huang is supported by the National Key R$\&$D Program of China (Grant no. 2020YFC2200903), the Preresearch Project on Civil Aerospace Technologies funded by China National Space Administration (Grant No. D010105), the Strategic Priority Research Program of Chinese Academy of Sciences (Grant no. XDA30040500).
%% Bibliography
%% Author year style
\bibliographystyle{plain}
\bibliography{refs}
\end{document}